\documentclass[preprints,article,accept,pdftex,moreauthors]{Definitions/mdpi} 

\usepackage[labelformat=simple]{subcaption}

\DeclareCaptionLabelFormat{subcaptionlabel}{\normalfont(\textbf{#2}\normalfont)}
\firstpage{2581} 
\pubvolume{23}
\issuenum{5}
\articlenumber{2581}
\pubyear{2023}
\copyrightyear{2023}
\externaleditor{Academic Editor: }
\datereceived{24 January 2023} 
\daterevised{16 February 2023}
\dateaccepted{21 February 2023} 
\datepublished{27 February 2023} 
\hreflink{https://doi.org/10.3390/s23052581}
\doinum{10.3390/s23052581}
\pdfoutput=1 

\Title{Improving BLE-Based Passive Human Sensing with Deep~Learning}

\TitleCitation{Iannizzotto, G.; Lo Bello, L.; Nucita, A. Improving BLE-Based Passive Human Sensing with Deep Learning. \textit{Sensors} \textbf{2023}, \textit{23}, 2581. \url{https://doi.org/10.3390/s23052581}}

\Author{{Giancarlo} Iannizzotto $^{1,}$*\orcidA{}, Lucia Lo Bello $^{2}$\orcidB{} and Andrea Nucita {$^{1}$}\orcidD{}}

\AuthorNames{Giancarlo Iannizzotto, Lucia Lo Bello and Andrea Nucita}

\AuthorCitation{{Iannizzotto, G.;} Lo Bello, L.; Nucita, A. }

\address{%
$^{1}$ \quad Department of Cognitive Sciences, Psychology, Education and Cultural Studies (COSPECS), \mbox{University of Messina}, 98122 Messina, Italy; {(A. Nucita):anucita@unime.it; (G. Iannizzotto):ianni@unime.it}\\
$^{2}$ \quad Department of Electrical, Electronic and Computer Engineering (DIEEI), University of Catania, \mbox{95125 Catania, Italy}; lucia.lobello@unict.it\\
}

\corres{Correspondence: ianni@unime.it}

\abstract{Passive Human Sensing (PHS) is an approach to collecting data on human presence, motion or activities that does not require the sensed human to carry devices or participate actively in the sensing process. 
In the literature, PHS is generally performed by exploiting the Channel State Information variations of dedicated WiFi, affected by human bodies obstructing the WiFi signal propagation path. However, the adoption of WiFi for PHS has some drawbacks, related to power consumption, large-scale deployment costs and interference with other networks in nearby areas. Bluetooth technology and, in particular, its low-energy version Bluetooth Low Energy (BLE), represents a valid candidate solution to the drawbacks of WiFi, thanks to its Adaptive Frequency Hopping (AFH) mechanism. This work proposes the application of a Deep Convolutional Neural Network (DNN) to improve the analysis and classification of the BLE signal deformations for PHS using commercial standard BLE devices. The proposed approach was applied to reliably detect the presence of human occupants in a large and articulated room with only a few transmitters and receivers and in conditions where the occupants do not directly occlude the Line of Sight between transmitters and receivers. This paper shows that the proposed approach significantly outperforms the most accurate technique found in the literature when applied to the same experimental data.}


\begin{document}

\section{Introduction}\label{Sec:Introduction}
Passive Human Sensing (PHS), often referred to as deviceless or device-free human sensing, is a technique to gather data on human presence, motion or activities that does not entail for the sensed human the need to carry devices or play an active role in the sensing process~\cite{Youssef2007}. PHS exploits the deformation of an RF signal emitted and received by the sensing infrastructure to collect information about the environment, the presence of human bodies in the area illuminated by the signal, and its reflected and refracted components~\cite{Woyach2006}. As a consequence, PHS differs radically from most indoor and outdoor localization approaches, which instead require the human to carry a handheld or wearable device that receives or transmits a localization signal to a support infrastructure~\cite{s21238086,Carvalho20181}. PHS also differs from vision-based human sensing~\cite{s19194129, IANNIZZOTTO2021115447, Aggarwal-Ryoo2011, Iannizzotto20113151} for several reasons. Firstly, PHS does not require Line of Sight (LoS) visibility between the sensor and the sensed human. Second, PHS does not suffer from low illumination quality. Finally, PHS does not raise the well-known privacy issues related to camera-based sensing~\cite{HUSSAIN2020102738, Iannizzotto2020,7412751, Battaglia2014}. 

Among the RF signals sources available for PHS, WiFi was soon identified as the most suitable one thanks to its wide diffusion and the flexibility of its devices, thus attracting the interest of researchers from the very beginning~\cite{Youssef2007, 7040509}. Early studies mainly focused on the modifications introduced in the Received Signal Strength Indicator (RSSI) of the WiFi signal by an obstructing human body. However, for about a decade, the Channel State Information (CSI) has been considered a more reliable and richer source of information~\cite{9613686}. In fact, RSSI is a scalar measure that characterizes the radio signal attenuation during propagation, therefore the sensing approaches based on RSSI alone (used either as a measure of path loss or as a radio-based fingerprint) suffer from dramatic performance degradation in complex situations due to multipath fading and changes in the environment~\cite{YangZhouLiu2013}. The Channel State Information instead offers a
set of channel measurements that describe the amplitudes and phases of each subcarrier of an Orthogonal Frequency Division Multiplexing (OFDM) transmission, which is typical of recent 802.11 a/g/n standards. RSSI is the superimposition of multipath signals with fast-changing phases, whereas CSI produces fine-grained, per-subcarrier information, thus allowing for a better discrimination of multipath signals. In the last decade, the analysis of the CSI variations enabled the experimentation of a large number of applications, such as human detection and counting, person identification/authentication, person localization and tracking, recognition of human activity and the detection and monitoring of vital signs~\cite{Yongsen2019,SOTO202299}.

Despite the wide interest raised in the research community, WiFi-based PHS has a number of drawbacks. First of all, the WiFi protocols do not support the coexistence of communication and sensing. As a consequence, trying to perform sensing and communication at the same time on the same network or using two networks working in proximity to each other can result in severe performance and accuracy degradation for both~\cite{9814523}, unless suitable mechanisms are introduced to support multiple traffic classes with different Quality of Service (QoS), thus giving higher priority to the sensing traffic~\cite{s20185195,sched2020}. This problem is even worsened by the 300-meter range of WiFi, which generates a relatively large collision domain for the two networks. Furthermore, WiFi is not designed for low power consumption. Consequently, since PHS requires frequent transmissions to allow prompt detection and adequate sampling of the sensed entity dynamics, the deployment of a WiFi-based sensing network requires careful planning to provide the devices with an adequate power source. Finally, large-scale deployments of WiFi devices, e.g., to monitor entire buildings or industrial plants~\cite{Huang2019}, can be quite expensive. 

 Bluetooth technology, in particular after the introduction of the Bluetooth Low Energy (BLE) specifications,
represents a valid alternative to WiFi. In fact, BLE~\cite{s22093523}:

\begin{itemize}
\item Is capable of working in noisy environments and in proximity to other wireless communication technologies, such as WiFi, thanks to its Adaptive Frequency Hopping (AFH) mechanism~\cite{FHSS} 
\item is integrated into most of  portable devices (such as tablets, smartphones, PDAs, etc.);
\item Is energy efficient; 
\item Allows for simple and flexible deployment in business, industrial and home environments, as BLE devices are small, minimally invasive and less expensive than other~solutions;
\item Provides an indoor communication range of approximately 20--30 m, depending on the specific device and the characteristics of the environment~\cite{s17020372}. As a consequence, the BLE probability of interfering with other networks working in proximity is significantly lower.
\end{itemize}

As reported in our recent survey on Bluetooth-based PHS~\cite{s22093523}, despite the evident advantages {and its wide adoption for active mobile localization}~\cite{9296763}---{that is, the localization of a moving transmitter with respect to a network of fixed receivers---}not many research works in the literature focus on the application of BLE to PHS. This is mainly due to the following three limitations of the BLE standard:
\begin{itemize}
    \item It is difficult to obtain RSSI samples at high rates from BLE-based devices; therefore, the ability of a BLE-based PHS system to match the dynamics of human activities and gestures may be insufficient for some applications~\cite{s21155181};
    \item The BLE protocol does not allow the application layer to know the current transmission frequency, that is, the PHS application has no legal (i.e., compliant with the protocol) way to extract the frequency selected by the AFH mechanism to transmit a specific message. This is a security-related constraint, so it will not be relaxed anytime soon;
    \item The WiFi CSI can be obtained directly from some devices through the WiFi multicarrier encoding mechanism, whereas BLE does not natively support such a measurement. 
\end{itemize}

Notwithstanding the limitations mentioned above, activities such as passive human detection~\cite{Munch2019}, counting~\cite{Munch20191}, and motion tracking~\cite{Brockmann2018} can be effectively performed using BLE networks, although with a generally lower accuracy than their WiFi counterparts~\cite{MAIS2017}. 
Despite the recent advancement and wide diffusion of Deep Learning (DL) techniques, only a few research works in the literature apply DL to BLE-based PHS.
In~\cite{jsan10020035}, a classroom scenario is considered and a Radial Basis Function Neural Network is used to classify RSSI samples emitted by carefully placed BLE beacons and received by a BLE receiver, with a varying number of occupants in the room. Unfortunately, the dataset used for the reported experiments is not publicly available and the paper does not report sufficient information to repeat the experiments. In~\cite{Billah_Saoda_Gao_Campbell2021} a rather complex reinforcement learning approach is proposed, using BLE RSSI for PHS and exploiting {$CO_2$} sensors and other IoT devices to generate feedback information for the reinforcement learning process. The presented approach is claimed to be adaptive to changes in the environment. However, the approach requires an initial fine-tuning process that closely resembles training and the presented results are comparable to those obtained with much simpler statistical approaches, such as the one in~\cite{Munch2019}. In~\cite{Demrozi2021EstimatingIO}, a BLE-based PHS is presented and compared with state-of-the-art WiFi-based competitors. However, the presented architecture  requires manual measurement of the distance between transmitters and receivers at training time. In addition, the best performance is obtained by forcing the sampling rate of the sensing network to very high rates (up to 200 Hz), i.e., well beyond the advertisement message transmission rate of standard BLE beacons.

This work proposes the application of a Deep Neural Network (DNN) to improve the analysis and classification of BLE signal deformations for Passive Human Sensing. Our aim was to improve over the current state-of-the-art in terms of both accuracy and flexibility. To this end, the proposed approach was applied to reliably detect the presence of human occupants in a large, articulated room with only a few BLE 4.x standard beacons as transmitters and a few receivers equipped with BLE 4.x standard modules. Furthermore, no specific restrictions were imposed on the location and motion dynamics of the room occupants. 

The main contributions of this work are:
\begin{enumerate}
    \item A PHS architecture for human detection based on standard BLE 4.x Commercial-Off-The-Shelf (COTS) devices that does not impose strong restrictions, such as direct Line-Of-Sight (LoS) visibility, on the position of the sensed occupants in the monitored environment. The transmitting devices are common BLE stand-alone beacons, which do not require any additional communication connection to external computational resources and can be battery-operated. The receiving device can be built on Arduino Zero or Raspberry Pi cards, equipped with their BLE onboard adapter;  
    \item A Deep Convolutional Neural Network to analyze the sequence of RSSI samples captured by the BLE receivers and extract the occupancy information of the area covered by the BLE sensing network. Our approach is novel compared to previous work, which adopted Long Short-term Memory (LSTM) networks~\cite{Billah_Saoda_Gao_Campbell2021}; 
    \item An assessment of the proposed architecture in a real-life environment, that is, a student laboratory where occupants are free to move in and out. The room was not rectangular but L-shaped with several non-LoS regions with respect to the BLE receivers, and the environment was polluted by RF noise due to several WiFi and Bluetooth operating devices that were randomly moved in and out of the room by the students;
    \item A comparison of the proposed approach with the one in the literature that offers the best performance~\cite{Munch2019}.
\end{enumerate}

\textls[-15]{The paper is organized as follows. Section \ref{Sec:MaterialsAndMethods} describes the proposed sensing architecture and experimental setup. Section \ref{Sec:Results} presents the experimental data obtained during the validation experiments and a comparison of our approach with the most accurate state-of-the-art approach from the relevant literature. Section \ref{Sec:Discussion} reports our analysis of the experimental results and Section \ref{Sec:Conclusion} presents our conclusions and suggestions for further~research.}

\section{Materials and Methods}\label{Sec:MaterialsAndMethods}
\begin{figure}[H]
{\captionsetup{position=bottom,justification=centering}
\begin{subfigure}{.49\columnwidth}
\vspace{0.22in}
  \centering
  \includegraphics[width=\columnwidth]{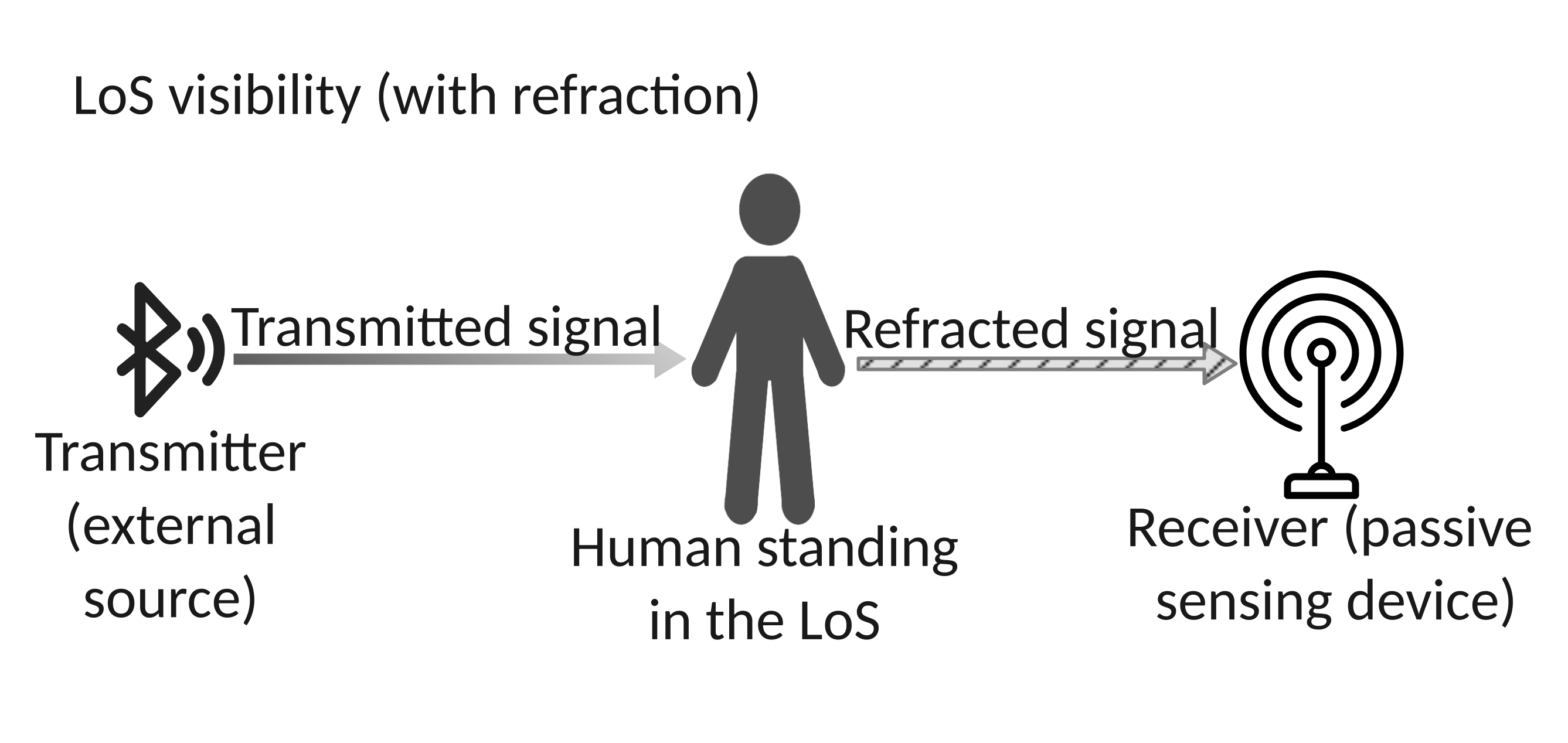}
  \caption{}
  \label{fig:fig01}
\end{subfigure}
\begin{subfigure}{.49\columnwidth}
  \centering
  \includegraphics[width=\columnwidth]{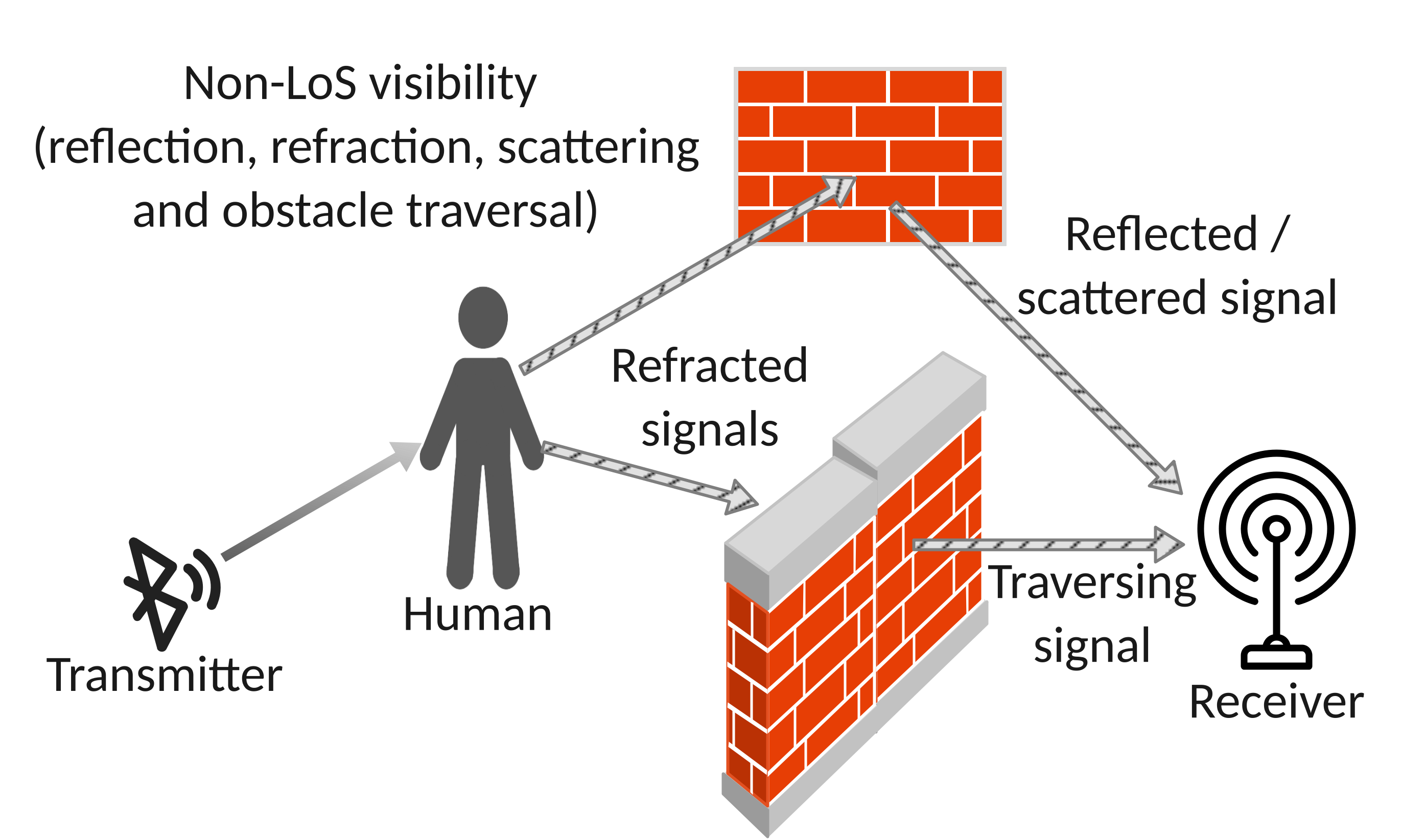}
  \caption{}
  \label{fig:fig02ab}
\end{subfigure}}
\vspace{0.1in}
\caption{{Line-of-Sight (LoS) visibility} (\textbf{a}). Non-Line-of-Sight and through-the-wall visibility (\textbf{b}). This figure was adapted from~\cite{s22093523}.}
\label{fig:longfig01}
\end{figure}

BLE-based PHS takes advantage of the deformations inducted in the RF signal by the human bodies that obtrude its pathway. When the obtruding body is situated exactly on a linear and clear (that is, otherwise unobtruded) pathway between the transmitter and the receiver, it affects the Line-of-Sight (LoS) visibility between the two devices and the signal that reaches the receiver is mainly the refracted component of the original signal (see Figure \ref{fig:longfig01}a). In general, the refracted component of the signal is deformed while traversing the human body and such a deformation transports useful information about the obtruding entity. Instead, if the obtruding body is not situated along a straight and clean pathway between transmitter and receiver, then the signal that reaches the receiver is refracted or reflected, either before or after traversing the obtruding body (Non-Line-Of-Sight visibility, see Figure \ref{fig:longfig01}b).\vspace{-9pt} 

\begin{figure}[H]
  \includegraphics[width=.95\columnwidth]{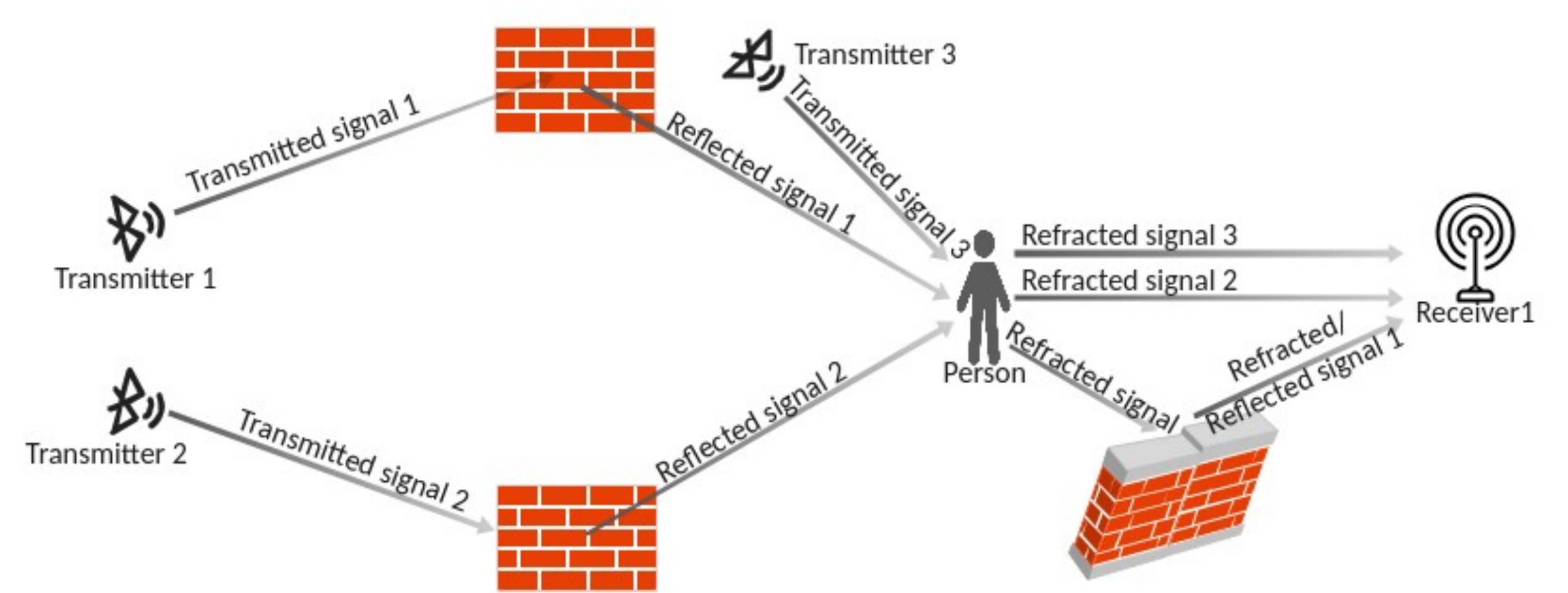}
\caption{Multipath (Non-LoS) PHS. Note that multiple signals traverse the same obtruding human body from different angles and with different time (phase) shifts, thus transporting spatially and temporally diverse information.}\label{fig:Drawing08}
\end{figure}

Since the exact position of each obtruding body is not known in advance and is not fixed, in order to reduce the number of transmitters and receivers, for the sensing process we relied on Non-Line-Of-Sight visibility. An interesting side effect of this approach is that multiple signals from different transmitters traverse the body and reach the receiver, thus transporting information acquired from different angles. Moreover, such signals go through different paths of different lengths; therefore, they have different phase shifts, i.e., they carry information on the body traversed at different instants in time~\cite{YangZhouLiu2013}. 
As a consequence, the sensing system acquires information at a much higher spatial and temporal resolution than LoS visibility would support (see Figure \ref{fig:Drawing08}). 

Clearly, the problem is how to extract this information. One reason is that, to benefit from the increased temporal resolution, accurate clock synchronization would be needed among all transmitters and receivers, but this feature is not supported by the BLE protocol. For the sake of simplicity, in this work we chose to treat the higher temporal resolution as redundancy, that is, we combined the samples acquired in very close time instants to produce more reliable average samples.

The BLE 4.x protocol only allows for the extraction of RSSI measurements from a message on the receiver side, that is, only a coarse-grained sampling of the received signal can be obtained. Moreover, security reasons impose that the AFH mechanism does not allow the application to know the channel from which the message was received. As RSSI has different mean values for each channel~\cite{7103024}, it is not possible to differentiate the RSSI samples based on the channel. This induces high variability in the RSSI distribution, which hinders the classification accuracy of most BLE-based PHS approaches. To overcome this problem, some works in the literature fix the transmission channel on the transmitter side~\cite{Brockmann2018}; however, such a trick is generally not allowed by COTS BLE 4.x beacons. A different solution proposed in the literature is to obtain the transmission channel by exploiting some specific vulnerability on the receiver side~\cite{Gentner2020IdentifyingTB}, but this possibility might disappear at any time with an OS update. 

As a consequence of the inability to determine the transmission channel of the received message, a very effective classification approach is needed that can take into account the large variability in the RSSI signal and produce an accurate classification.

\begin{figure}[H]
  \includegraphics[width=3.5in]{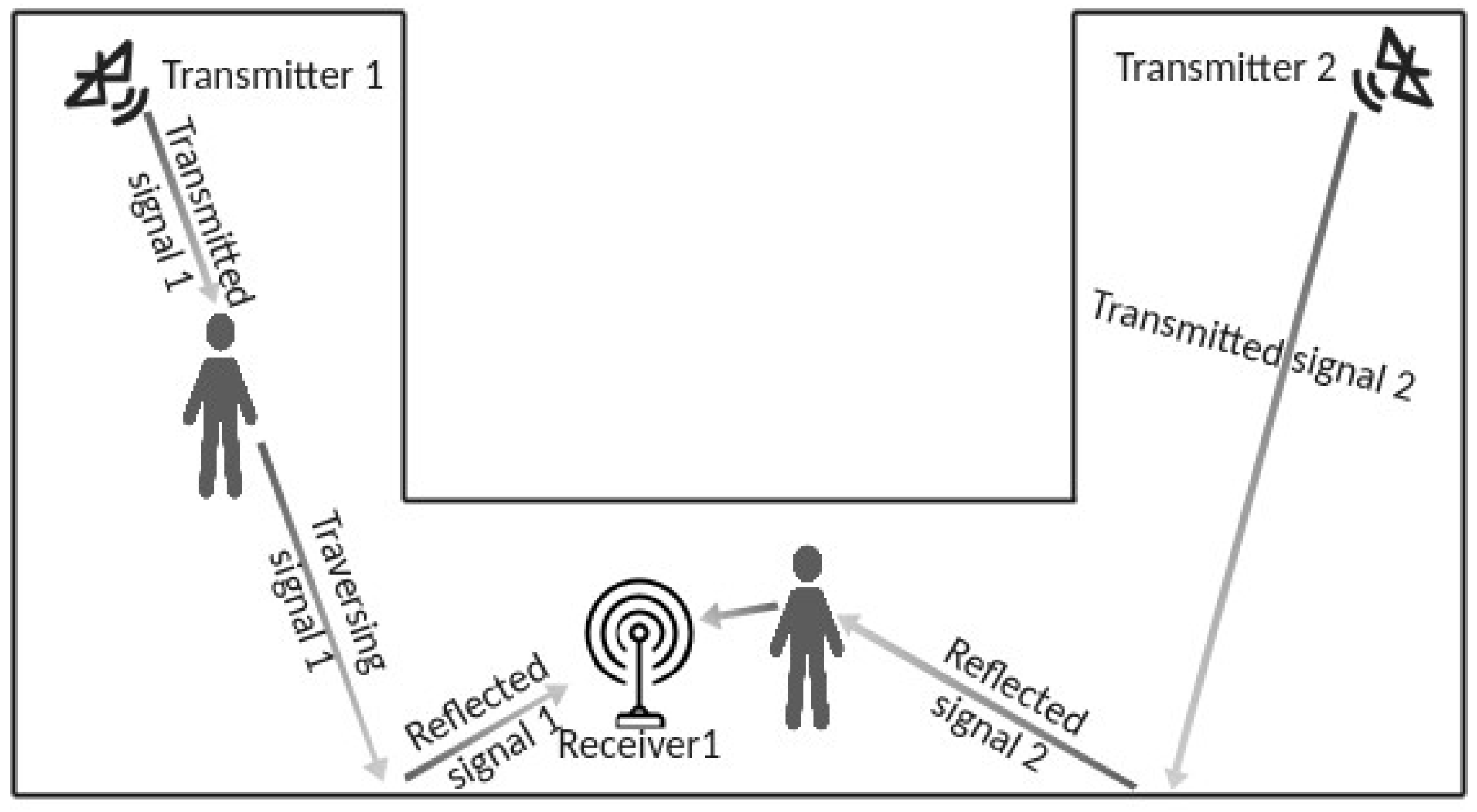}
\caption{{Constraints on the number and position of beacons and receivers. The beacons must be placed in such a way as to illuminate each occupant in the monitored room. The receivers must be placed in such a way that for each occupant, at least one transmission from a beacon traverses the occupant and is received by one of the receivers. The transmission may be reflected by some surface, i.e., it is not necessary that the occupant lays on the Line of Sight between the beacon and the receiver. }}\label{fig:beacons_and_receivers}
\end{figure}

In our architecture, the PHS network is based on $n$ BLE 4.x beacons and $m$ receiver devices equipped with BLE 4.x modules. The number $n$ of beacons is only constrained by the need to illuminate with the BLE signal all the occupants of the monitored area, wherever they are. The number $m$ is constrained by the need that, for each occupant, at least one receiver obtains from one of the BLE beacons of the sensing network an advertising message that has traversed the body of the occupant. {The advertising message may have been reflected by any surface before reaching the receiver, i.e., the occupant does not need to occlude the Line of Sight between the beacon and the receiver, as illustrated in Figure} \ref{fig:beacons_and_receivers}.

\begin{figure}[H]
  \includegraphics[width=\columnwidth]{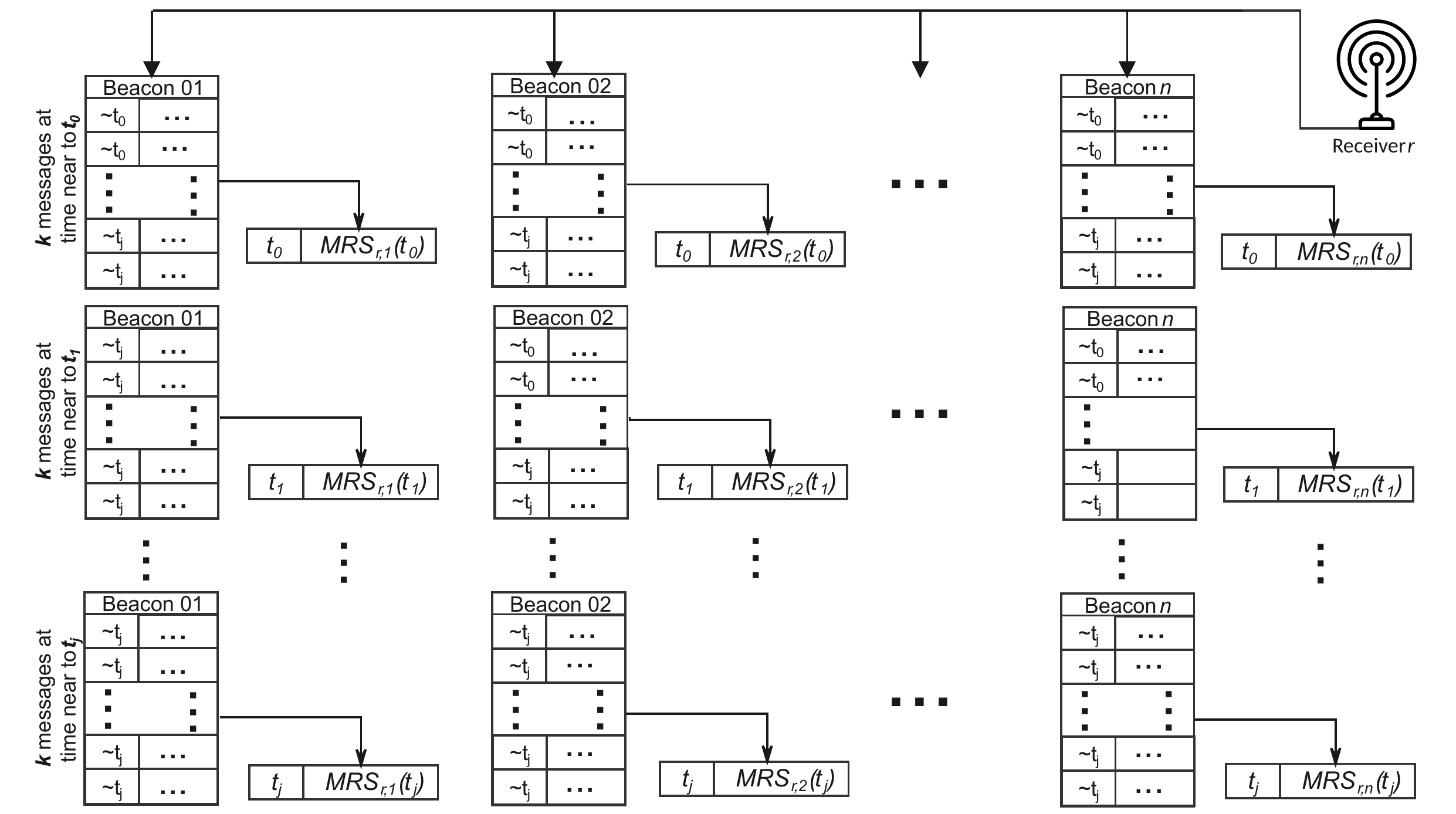}
\caption{{Graph of data collection and subsampling in a generic receiver $r$. The samples sent by each beacon are grouped in chunks of $k$, a representative sample is selected as the median element of the $k$ samples and keeps its timestamp (see text). The symbol $\mathtt{\sim}t_j$ means ``approximately $t_j$'', that is, all samples from the same group of $k$ have similar timestamps. }}\label{fig:system_schema}
\end{figure}

As the BLE devices are not synchronized, we cannot exploit the phase shift of different messages sent at the same time, for example, to improve spatial resolution~\cite{5646822}. Consequently, we prefer to sacrifice temporal resolution for noise robustness and combine the RSSI of $k$ messages, sent by the same $i$-th beacon {and received by receiver $r$} with similar timestamps around the time instant $t$, into a single representative sample calculated as the median of the {$k$} RSSI values, which we call the Median RSSI Sample {received by $r$} at time $t$ {($MRS_{r,i}(t)$)}. This approach strongly reduces the effects of unwanted abrupt changes in RSSI due to the AFH mechanism of the BLE protocol, while also reducing the temporal resolution (that is, the PHS sampling frequency) by a factor of $k$. 
{The temporal subsampling introduced above is illustrated in} Figure \ref{fig:system_schema}. 

At each receiver and at each receiving time instant $j$, the Median RSSI Samples with similar timestamps, but transmitted by different beacons, are collected in a row of RSSI samples marked with the representative timestamp $j$. Thus, each receiver $r$ maintains a list $MRSL_r$ of rows, each identified by a representative time instant $j$ and containing the sequence of $n$ values $MRS_{r,i}(j)$ received at that time and sent by the different beacons.

{The $m$ lists (which we named $MRSL_r$) are all sent to a central server that combines them into a single list and orders its rows according to their representative timestamps. This list is the input for the classifier}. At this point of the algorithm the temporal resolution is sufficiently coarse to allow us to neglect the effects of loss of synchronization between the receivers and the central server. In fact, in general the transmission rate of the advertising messages in BLE 4.x is 100 ms, therefore the temporal resolution at this stage of the algorithm is $1/(k*0.1)$ samples per second. On the server, the list of rows is fed to a DNN that classifies them and produces a decision about whether there are human occupants in the monitored area or not.

\begin{figure}[H]
  \includegraphics[width=\columnwidth]{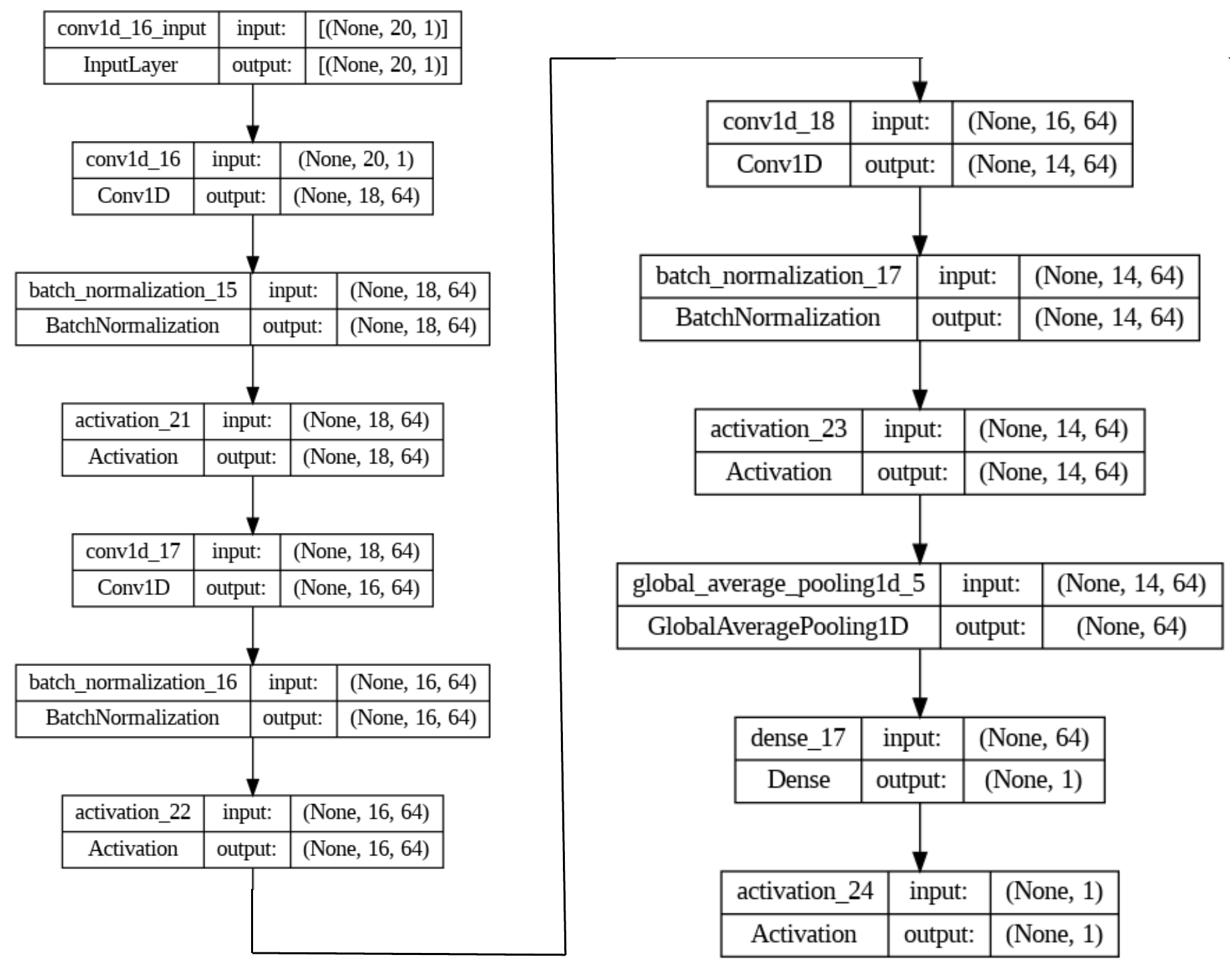}
\caption{The proposed DNN architecture with a total number of 25,793 parameters} \label{fig:DNN}
\end{figure}

{Several DNN architectures were investigated to build the classifier. In particular, a baseline network with three dense layers and dropout, a single-layer Long Short-Term Memory (LSTM) network, a two-layer LSTM network and a two-layer 1D convolutional network were initially tested. The proposed DNN (see Figure} \ref{fig:DNN}), {a pure convolutional network with 25,409 trainable parameters composed of three 1D convolutional layers with batch normalization, a global average pooling and a final dense output layer, consistently outperformed its competitors in all tests. For completeness, the two-layer LSTM architecture, the best performer among competitors of the proposed architecture, is included in the experiments reported in Section }\ref{Sec:Results}.

After deploying the BLE network, training and testing data are collected by manually annotating the time at which each occupant entered and exited the monitored area. This step is only needed to collect data for training and performance analysis. Once a sufficient number of samples are collected, the DNN can be trained. The Maximum Training Epochs parameter is set to 500, however in our experiments training generally stopped around 300~epochs as an effect of the early stopping mechanism~\cite{Song2019PrestoppingHD}.

\section{Results}\label{Sec:Results}
In our experiments, the BLE traffic was generated by COTS standalone and fully standard BLE 4.0 beacons, Eddystone-compliant and battery-operated. The receivers were built on consumer-grade Raspberry Pi 3B embedded computers, equipped with a BLE 4.1-compliant chipset onboard. A commercial notebook equipped with an Intel i7-10750H CPU @ 2.60~GHz, 16 GB RAM main memory, and an NVIDIA GTX 1650Ti GPU was used to both train and run the Deep Neural Network adopted for the classification of the BLE signal RSSI samples {into the ``Presence'' and ``Non-Presence'' classes}. With the proposed server architecture, the training was completed in a few minutes, thus allowing a very simple deployment of the PHS~system. 

\begin{figure}[H]
\centering
  \includegraphics[width=\columnwidth]{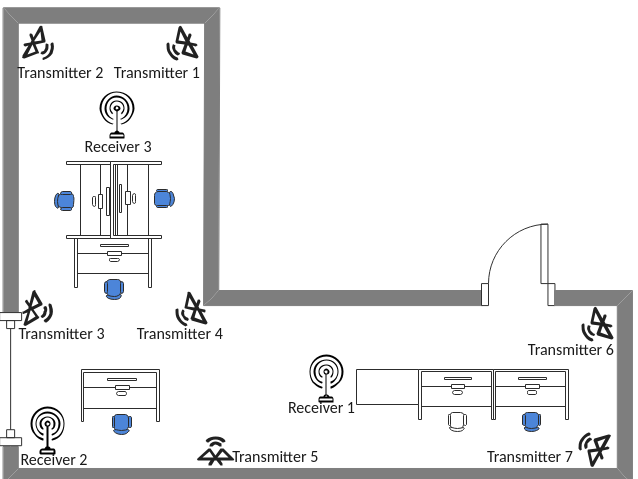}
\caption{Arrangement of BLE devices in the laboratory testbed. The blue chair symbols represent the average positions of the occupants, i.e., the white chair was empty most of the time.}\label{fig:Map}
\end{figure}

The room where the experiments were performed was a research laboratory with six work stations and several computers and other instruments. As the room was L-shaped, we placed approximately one beacon at each corner to fully illuminate the room area. Similarly, we placed three receivers in such a way that most of the beacon signals would be consistently received. No specific optimizations were performed during the placement of the transmitting and receiving devices. In our experience, the exact positioning of the beacons and receivers is not critical; however, we tried not to place the receivers too close to the beacons. Figure \ref{fig:Map} shows a map of the laboratory with the location of the transmitters, receivers, and work stations that were used by the occupants during the experiment. Noticeably, the occupants occasionally moved across the room as in normal laboratory operations; {however, the number of moving occupants was not logged in time, so the only information available is that the occupants were not forced to sit at their workstation, but could stand up and move freely across the room}.

\begin{figure}[H]
{\captionsetup{position=bottom,justification=centering}
\begin{subfigure}{.49\columnwidth}
  \centering
  \includegraphics[width=\columnwidth]{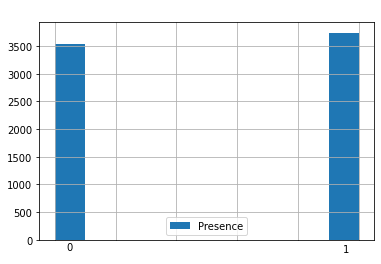}
  \caption{Training and validation dataset}
  \label{fig:figbalance01}
\end{subfigure}
\hfill
\begin{subfigure}{.49\columnwidth}
  \centering
  \includegraphics[width=\columnwidth]{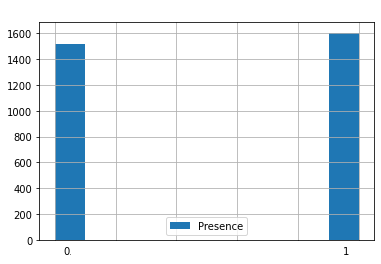}
  \caption{Test dataset}
  \label{fig:figbalance02}
\end{subfigure}}
\vspace{0.1in}
\caption{Histograms of the ``Presence'' values of the samples, showing the balance between the Presence and Non-Presence samples in the Training (\textbf{a}) and Test (\textbf{b}) datasets.}
\label{fig:longfigbalance}
\end{figure}

For our experiments, after cleanup we obtained a total of {10,412} samples from $r=3$ receivers and $b=7$ beacons during several hours of daily working and non-working time. We randomly split the samples into 7288 (i.e., two-thirds of the total) for training {and validation}, and 3124 (i.e., one-third of the total) for testing. The two datasets are approximately balanced, that is, in both datasets, the numbers of samples belonging to the ``Presence'' and ``Non-presence'' classes {differ by less than 10\%}, as shown in Figure \ref{fig:longfigbalance}. {The Test dataset was then put aside for the final evaluation, while the 7288 samples were further equally split into the ``Training'' and ``Validation'' datasets}. 

\vspace{-6pt}

\begin{figure}[H]
{\captionsetup{position=bottom,justification=centering}
\begin{subfigure}{.49\columnwidth}
\vspace{0.07in}
  \centering
  \includegraphics[width=\columnwidth]{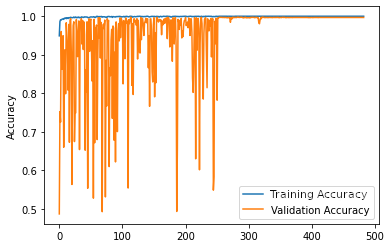}
  \caption{}
  \label{fig:figaccuracy01}
\end{subfigure}
\hfill
\begin{subfigure}{.49\columnwidth}
  \centering
  \includegraphics[width=\columnwidth]{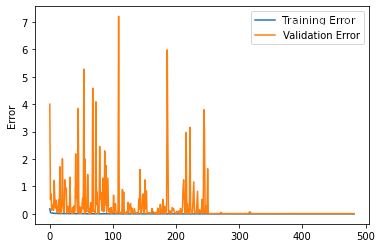}
  \caption{}
  \label{fig:figaccuracy02}
\end{subfigure}}
\vspace{0.1in}
\caption{Graph showing Accuracy and Validation Accuracy (\textbf{a}) and Error and Validation Error (\textbf{b})~during the training process.}
\label{fig:longfigaccuracy}
\end{figure}

After an initial oscillation phase, the training process steadily converges towards an accuracy of 100\%  and a 0\% error in both training and validation, as shown in Figure \ref{fig:longfigaccuracy}.

\begin{figure}[H]
  \includegraphics[width=3.5in]{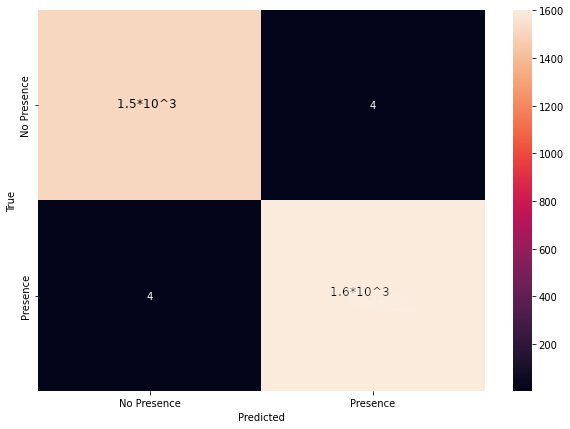}
\caption{{The confusion matrix relevant to the presented experiments.}} 
\label{fig:confusion}
\end{figure}

After training and validating the proposed model, we tested it on the Test dataset, made of samples that did not belong to the Training and Validation datasets and were, therefore, completely unknown to the model. The discrimination ability of the model is well evidenced by the confusion matrix related to our experiments, shown in Figure \ref{fig:confusion}.

The accuracy obtained during the tests, i.e., 0.9974 (99.74\%), was consistently achieved and even exceeded along several iterations of the training--validation--test sequence.

To compare our architecture with the literature, we selected the state-of-the-art approach that a recent survey~\cite{s22093523} reported to be the best performer in a similar experimental setup~\cite{Munch2019} and tested it with our datasets. In~\cite{Munch2019}, a network of BLE beacons and BLE receivers was used to sense the occupancy of a lecture room. Several approaches for data preprocessing and several classifiers were investigated and evaluated and a final 98.97\% accuracy is reported for the best combination. We re-trained and tested on our datasets the same data preprocessing and classification methods described in~\cite{Munch2019} and selected the best performing combination, that is, Support Vector Machine (SVM) with linear kernel (SVM-LK) classifier in combination with the DS3 data preprocessing method. The results obtained are shown in Table \ref{tab:metrics} for comparison with our approach {and with the two-layer LSTM network that performed second during our preliminary investigation (see Section }\ref{Sec:MaterialsAndMethods}).

\begin{table}[H] 
\caption{Performance comparison of our approach with the method proposed in~\cite{Munch2019}.\label{tab:metrics}}
\newcolumntype{C}{>{\centering\arraybackslash}X}
\begin{tabularx}{\textwidth}{CCCCC}
\toprule
\textbf{Approach} & \textbf{Accuracy} & \textbf{Precision} & \textbf{Recall} & \textbf{F1-Score}\\
\midrule
SVM-LK		            & 0.9682    & 1.00  & 0.94  & 0.97\\
{LSTM}                    & 0.9819    & 0.98  & 0.98  & 0.98\\ 
\textbf{{our approach}}	& \textbf{0.9974}	& \textbf{1.00}  & \textbf{1.00}  & \textbf{1.00}\\
\bottomrule
\end{tabularx}
\end{table}

As a further test, to investigate the robustness of our approach with respect to the number and location of the transmitting and receiving devices, we progressively dropped the receivers and beacons and measured the performance of the degraded sensing system. 
With reference to Figure \ref{fig:Map}, Table \ref{tab:impairedmetrics} shows the degraded network configurations and the performance obtained with our approach and with the method reported in~\cite{Munch2019}.

\begin{table}[H] 
\caption{Performance comparison of our approach with the method proposed in~\cite{Munch2019} {and with the LSTM network}, in ``degraded network conditions'' (dropped receivers and/or beacons). The first column indicates which devices were available in each degraded configuration (i.e., the devices that are not indicated were dropped). r1, r2, r3 are the receivers and b1,$\ldots$, b7 are the beacons.\label{tab:impairedmetrics}}
\newcolumntype{C}{>{\centering\arraybackslash}X}
\begin{tabularx}{\textwidth}{CC C C C C}
\toprule
\textbf{Avail. Devs.} & \textbf{Approach} & \textbf{Accuracy} & \textbf{Precision} & \textbf{Recall} & \textbf{F1-Score}\\
\midrule
\multirow{3}{2.2cm}{\centering All beacons, \\ r1 \& r3} & SVM-LK & 0.6508 & {0.74}  & {0.46}  & {0.57}\\
& {LSTM}	& 0.9771	& 0.98  & 0.98  & 0.98\\
& \mbox{\textbf{{our~approach}}}	& \textbf{0.9886}	& \textbf{0.99}  & \textbf{0.99}  & \textbf{0.99}\\
\midrule
\multirow{3}{2.2cm}{\centering All beacons, \\ only r2} & SVM-LK & {0.7698} & {0.93}  & {0.59}  & {0.72}\\
& {LSTM}	& 0.9635	& 0.96  & 0.96  & 0.96\\
& \textbf{our~approach}	& \textbf{0.9713}	& \textbf{0.97}  & \textbf{0.97}  & \textbf{0.97}\\
\midrule
\multirow{3}{2.2cm}{\centering All beacons, \\ only r1} & SVM-LK & {0.6905} & {0.79}  & {0.52}  & {0.63}\\
& {LSTM}	& 0.9615	& 0.96  & 0.96  & 0.96\\
& \textbf{our~approach}	& \textbf{0.9713}	& \textbf{0.97}  & \textbf{0.97}  & \textbf{0.97}\\
\midrule
\multirow{3}{2.2cm}{\centering b1, b2, b6, b7, \\ r1 \& r3} & SVM-LK & {0.5556} & {0.57}  & {0.48}  & {0.52}\\
& {LSTM}	& 0.9559	& 0.96  & 0.95  & 0.96\\
& \textbf{our~approach}	& \textbf{0.9669}	& \textbf{0.97}  & \textbf{0.97}  & \textbf{0.97}\\
\bottomrule
\end{tabularx}
\end{table}

\section{Discussion}\label{Sec:Discussion}

To effectively test our approach, we selected a very general scenario in terms of both spatial configuration and usage patterns. In particular: 
\begin{itemize}
    \item The shape of the laboratory is concave, so some receiver/beacon couples are not in direct LoS; 
    \item The laboratory is used for research and not for lecturing, so the number of occupants and the entrance and exit times are not fixed;
    \item The number of occupants is generally small, i.e., from zero to five, therefore the RSSI variation due to the presence of occupants in the laboratory can be very small; 
    \item Computers and other transmitting devices pollute the BLE transmission frequencies, thus triggering the BLE Frequency Hopping mechanism frequently;
    \item The beacons and the receivers were positioned according to general illumination considerations and not taking into account the expected location of the occupants. 
\end{itemize}

Conversely, the scenarios adopted in other works in the literature, including the one compared with our method in Section \ref{Sec:Results}, are generally nearly ideal, with occupants restricted within specific areas for most of the time and beacons placed within the LoS of the receivers in such a way that any occupant would fully obstruct the LoS.

Our approach achieved an accuracy of 99.74\% with 100\% f1-score. To the best of our knowledge, this is the best performance in the literature considering the test scenario. Moreover, our approach performed considerably better (+3\% accuracy and f1-score) than its direct competitor, as reported in Table \ref{tab:metrics}.

In addition, we demonstrated that our approach does not require direct LoS visibility between beacons and receivers and does not need a large number of beacons and receivers to achieve state-of-the-art performance. Table \ref{tab:impairedmetrics} shows that dropping half of the beacons or two out of three receivers, thus breaking the LoS visibility between most of the beacon/receiver couples, only marginally reduces the detection accuracy.

Finally, the proposed detection architecture is lightweight enough to run on an embedded computer such as a Raspberry Pi 3B+. In fact, the bare RSSI measurement task can be easily performed by much cheaper and less power-intensive hardware, such as the Raspberry Pi0. Instead, we purposely chose to build the receivers on the more powerful Pi3B+ platform because in future work we plan to run the detection software on each receiver and develop a distributed sensing platform instead of delegating all the computation to a centralized server.

Two important limitations affect the proposed approach, the first being its intrinsically low detection speed, due to the low sampling rate of its architecture. Based on the analysis of RSSI of advertising messages transmitted by commercially available BLE 4.x beacons, the sampling rate of the proposed architecture is bound by the maximum advertising rate of the BLE 4.x standard, that is, 50 messages per second. Moreover, due to the subsampling mechanism introduced to deal with the lack of inter-node synchronization, the sampling rate is further divided by $k$. In our experiments, the actual sampling rate was reduced to 2~Hz, which is sufficient in most cases but not for higher dynamics applications, for example, when a person rapidly traversing a small area needs to be detected.

Consequently, as a further development of our PHS platform, we plan to exploit the enhanced services and features of the BLE 5.x architecture~\cite{s22072759,s21113589} to improve the current spatial and time resolution of our PHS platform, thus also supporting the estimation of the number of occupants in the monitored area and allowing for the recognition of their~activity. 

A second limitation, shared with most other approaches in the literature, is that a training phase is needed to calibrate the PHS before it can be used effectively to detect the presence of occupants in a room. This procedure is performed only once at deployment time, takes about 3--4 h, and requires manual recording of the presence and absence of occupants in the monitored area. 

\section{Conclusions}\label{Sec:Conclusion} 

 This work presented a novel architecture for passive human sensing based on a network consisting of commercial BLE 4.x beacons and receivers built with a Raspberry Pi board equipped with its standard on-board BLE 4.x module. A Deep Convolutional Neural Network (DNN) was used to analyze and classify the RSSI samples of the signals emitted by the BLE beacons and extract the occupancy information of the area covered by the BLE sensing network. 

 The advantages and limitations of the architecture presented were illustrated and a detailed description of the experimental results conducted to assess its validity was~provided.
 
 In our experiments, the proposed approach outperformed the state-of-the-art method described in~\cite{Munch2019} and achieved an accuracy of 99.74\% and a 100\% f1-score in a fairly general scenario, thus was demonstrated to be, to our knowledge, the best performer with respect to the literature. 

\vspace{6pt} 

\authorcontributions{{Conceptualization, G.I. and A.N.; Methodology, G.I., L.L.B. and A.N.; Validation, L.L.B. and A.N.; Investigation, G.I.; Writing---original draft, G.I.; Writing---review \& editing, L.L.B. All authors have read and agreed to the published version of the manuscript.}}

\funding{This research received no external funding.}

\institutionalreview{Not applicable.}

\informedconsent{Not applicable.}

\dataavailability{Data available upon request from corresponding author.} 

\conflictsofinterest{The authors declare no conflict of interest.} 

\begin{adjustwidth}{-\extralength}{0cm}

\reftitle{References}

\PublishersNote{}
\end{adjustwidth}
\end{document}